# Gravitational field of domain wall in Lyra geometry

[1]F. Rahaman and P. Ghosh

Department of Mathematics, Jadavpur University, Kolkata-700032, India.

**[1]**E- Mail: farook_rahaman@yahoo.com

## Abstract:

In this paper, we study the domain wall with time dependent displacement vectors based on Lyra geometry in normal gauge i.e. displacement vector $\phi^*_i = [\beta(t), 0, 0, 0]$. The field theoretic energy momentum tensor is considered with zero pressure perpendicular to the wall. We find an exact solutions of Einstein's equation for a scalar field $\varphi$ with a potential $V(\varphi)$ describing the gravitational field of a plane symmetric domain wall. We have seen that the hyper surfaces parallel to the wall ( $z$ = constant ) are three dimensional de-sitter spaces. It is also shown that the gravitational field experienced by test particle is attractive.

## Introduction:

The origin of the structure in the Universe is one of the greatest cosmological mysteries even today. Extended topological objects such as monopoles, strings and domain walls may play a fundamental role in the formation of our Universe. In recent , Pando,Valls-Gabaud and Fang [1] have proposed that the topological defects are responsible for structure formation of our Universe. The existence of domain wall is associated with breaking of a discrete symmetry i.e. the vacuum manifold M consists of several disconnected components. So the homotopy group $\pi_0(M)$ is non trivial ( $\pi_0(M) \neq I$ ) [2].
 In recent years, due to the new scenario of galaxy formation as suggested by Hill, Schramm and Fry [3], the study of domain walls with finite thickness has gained renewed cosmological interest. As the domain walls are associated with a late time phase transitions, so the energy associated with the walls are very small in compare to GUT domain walls [2-7]. In 1989, Widrow [4] proved that a thick wall with non zero stress along a direction perpendicular to the plane of the wall does not allow static metric to be regular through out entire space.

The study of domain walls usually falls into two categories: One can study the field equations as well as equation of the domain wall treated as the self interacting scalar field. The second is to assume a certain form of energy stress tensors for a domain wall. And then solve the corresponding gravitational field equations. The first seems more complete. In 1990, Goetz [5] studied the gravitational field of plane symmetric thick domain wall by solving the coupled Einstein scalar field equations. He found that the metric is time dependent, where as the scalar field has a time independent Kink like shape. Recently, Rahaman et al have extended Goetz's work in spherically symmetric spacetime [5]. In 1993, Mukherji [6] rederived this solution. In 1994, Wang [7] studied the gravitational field of thick domain wall.

In last few decades, there has been considerable interest in Alternative theories of gravitation. The most important among them are scalar-tensor theories proposed by Lyra [8] and by Brans-Dicke[8]. Lyra [8] proposed a modification Riemannian geometry by introducing a gauge function in to the structure less manifold that bears a close resemblance to Weyl's geometry. In general relativity Einstein succeeded in geometrising gravitation by identifying the metric tensor with the gravitational potentials. In the scalar-tensor theory of Brans-Dicke, on the other hand, scalar field remains alien to the geometry. Lyra's geometry is more in keeping with the spirit of Einstein's principle of geometrisation, since both the scalar and tensor fields have more or less intrinsic geometrical significance.

In the consecutive investigations, Sen [9] and Sen and Dunn [10] proposed a new scalar tensor theory of gravitation and constructed an analog of the Einstein field equation based on Lyra's geometry which in normal gauge may be written as

$$R_{ik} - \tfrac{1}{2} g_{ik} R + (3/2)\phi^*_i \phi^*_k - \tfrac{3}{4} g_{ik} \phi^*_m \phi^{*m} = - T_{ik} \qquad \ldots(1)$$

where, $\phi^*_i$ is the displacement vector and other symbols have their usual meaning as in Riemannian geometry.

Halford [11] has pointed out that the constant displacement field $\phi^*_i$ in Lyra's geometry play the role of cosmological constant $\Lambda$ in the normal general relativistic treatment. According to Halford the present theory predicts the same effects within observational limits, as far as the classical solar system tests are concerned, as well as tests based on the linearised form of field equations. Soleng [12] has pointed out that the constant displacement field in Lyra's geometry will either include a creation field and be equal to Hoyle's creation field cosmology or contain a special vacuum field which together with the gauge vector term may be considered as a cosmological term.

Subsequent investigations were done by several authors in cosmology and topological defects within the framework of Lyra geometry [13]. Recently, Rahaman et al have done a work related to domain wall within the framework of Lyra geometry but with an ad hoc energy momentum tensor [14].

In this paper, we shall extend that work by considering a field theoretic energy momentum tensor. Our approach consists of working of out field equations in much the same way as Goetz did in solving the same problem in general relativity. Here we assume, the time like displacement vector, $\phi^*_i = [\ \beta(t), 0, 0, 0\ ]$.

We find a solution for an arbitrary potential $V(\varphi)$ with an arbitrary distribution of scalar field $\varphi$ whose metric represents the three dimensional de-sitter spaces in the z = constant hyper surfaces. If the domain walls exist in the present universe, they could be detected by their gravitational interactions with photons and other massive particles. The gravitational field is shown to be repulsive by studying the motion of test particle in the gravitational field of domain wall.

## 2. Basic Equations:

In this section, we develop a general relativistic model of a plane symmetric domain wall in Lyra geometry by considering field theoretic energy momentum tensor for a self-interaction scalar field. Here the energy momentum tensor describing a scalar field $\varphi$ with self-interaction contained in a potential $V(\varphi)$ takes the form [5]

$$T_{ab} = \tfrac{1}{2}\,\partial_a \varphi\, \partial_b \varphi - [\ \tfrac{1}{2}\,(\partial_k \varphi\, \partial^k \varphi) - V(\varphi)\ ]\ g_{ab} \qquad \ldots\ldots(2)$$

Here the energy stress components in co moving coordinates for thick wall with metric ansatz

$$ds^2 = e^A (dt^2 - dz^2) - e^C (dx^2 + dy^2) \qquad \ldots (3)$$

are given by

$$T_t^t = \tfrac{1}{2}\, e^{-A}\,(\varphi^1)^2 + \tfrac{1}{2}\, e^{-A}\,(\varphi^*)^2 + V(\varphi) \qquad \ldots(4)$$

$$T_z^z = -\tfrac{1}{2}\, e^{-A}\,(\varphi^1)^2 - \tfrac{1}{2}\, e^{-A}\,(\varphi^*)^2 + V(\varphi) \qquad \ldots(5)$$

$$T_x^x = T_y^y = \tfrac{1}{2}\, e^{-A}\,(\varphi^1)^2 - \tfrac{1}{2}\, e^{-A}\,(\varphi^*)^2 + V(\varphi) \qquad \ldots(6)$$

$$T_t^z = e^{-A}\, \varphi^1\, \varphi^* \qquad \ldots(7)$$

where A, C and $\varphi$ are functions of z and t.

The field equation (1) for metric (3) reduce to

$$\tfrac{1}{4}\, e^{-A}\,[\,-4\,C^{11} - 3\,(C^1)^2 + 2\,A^1 C^1\,] + \tfrac{1}{4}\, e^{-A}\,[\,(C^*)^2 + 2\,A^* C^*\,] - \tfrac{3}{4}\,\beta^2\, e^{-A}$$
$$= \tfrac{1}{2}\, e^{-A}\,(\varphi^1)^2 + \tfrac{1}{2}\, e^{-A}\,(\varphi^*)^2 + V(\varphi) \qquad \ldots(8)$$

$$\tfrac{1}{4}\, e^{-A}\,[\,4\,C^{**} + 3\,(C^*)^2 - 2\,A^* C^*\,] + \tfrac{1}{4}\, e^{-A}\,[\,-(C^1)^2 - 2\,A^1 C^1\,] + \tfrac{3}{4}\,\beta^2\, e^{-A}$$
$$= -\tfrac{1}{2}\, e^{-A}\,(\varphi^1)^2 - \tfrac{1}{2}\, e^{-A}\,(\varphi^*)^2 + V(\varphi) \qquad \ldots(9)$$

$$\tfrac{1}{4} e^{-A} [-2 A^{11} - 2 C^{11} - (C^1)^2] + \tfrac{1}{4} e^{-A} [2 C^{**} + 2 A^{**} + (C^*)^2] + \tfrac{3}{4} \beta^2 e^{-A}$$

$$= \tfrac{1}{2} e^{-A} (\varphi^1)^2 - \tfrac{1}{2} e^{-A} (\varphi^*)^2 + V(\varphi) \qquad \ldots(10)$$

$$\tfrac{1}{2} [-C^{*1} + C^*(A^1 - C^1) + A^1 C^*] = \varphi^1 \varphi^* \qquad ..(11)$$

Also the explicit expression for the scalar field equation for this metric is

$$e^{-A} [\varphi^{**} + \varphi^* C^*] - e^{-A} [\varphi^{11} + \varphi^1 C^1] = (dV/d\varphi) \qquad \ldots..(12)$$

[ '*' and '1' are differentiations w.r.t. 't' and 'z' respectively ]

## 3. Solutions:

Here we shall make an attempt to examine whether it is possible to find any solution satisfying both the Euler-Lagrange equation as well as Einstein equations.
Similar to Goetz [5] to solve the field equations we assume the following relation between energy stress components as

$$T_t^t = T_x^x = T_y^y \qquad \ldots..(13)$$

Here, $T_t^t$ is the energy density of the wall which is again equal to the tensions along x and y directions in the plane of the wall.
Following eq.(13), we get $\varphi^* = 0$ i.e. $\varphi$, the scalar field is static in nature.
Thus,

$$\varphi = \varphi(z) \qquad \ldots(14)$$

Now the scalar field equation (12) takes the form

$$[\varphi^{11} + \varphi^1 C^1] = -e^A (dV/d\varphi) \qquad \ldots..(15)$$

As the field equations are complicated, so to obtain the solutions, we assume the metric coefficients in separable form of z and t as

$$A = A_1(z) + A_2(t) \; ; \quad C = C_1(z) + C_2(t) \qquad \ldots..(16)$$

Since the coefficient of $\varphi^1$ term of eq.(15) is a function of z only, so the coefficient of potential term of eq.(15) must be function of z only:

Thus,
$$A = A_1(z) \qquad \ldots(17)$$

Moreover, $\varphi^* = 0$ implies $G_t^z = 0$.

Hence from eq.(11), we get

$$C_2^* ( A_1^1 - C_1^1 ) = 0 \qquad \ldots(18)$$

This implies either $C_2^* = 0$ or $A_1^1 = C_1^1$.

If $C_2^* = 0$, then the model becomes a static model, so we reject this case.

We choose,

$$A_1^1 = C_1^1 \text{ i.e. } A_1 = C_1 \qquad \ldots(19)$$

Similar to Goetz's approach, we assume, $T_z^z = 0$ and this implies,

$$V(\varphi) = \tfrac{1}{2} \exp(-A)\, \varphi^{1\,2} \qquad \ldots(20)$$

By adding eqs.(8) and (9) and using eqs.(16),(17),(19) and (20), we get

$$C_1^{11} + C_1^{1\,2} + \varphi^{1\,2} = C_2^{**} + C_2^{*\,2} = k \qquad \ldots(21)$$

Following the relation $(8) + (10) - 2.(9)$, we get

$$8\, C_1^{11} + 4\, (C_1^1)^2 + 8\, (\varphi^1)^2 = -6\, C_2^{**} - 4\, (C_2^*)^2 - 3\, \beta^2 = -n \qquad \ldots(22)$$

[ n and k are separation constants]

Considering eqs. (21) and (22), we get solutions for space part as

$$C_1 = C_0\, z \qquad \ldots(23)$$

$$\varphi_1 = \varphi_0\, z \qquad \ldots(24)$$

where, $C_0 = \sqrt{[(2/3)(k + \tfrac{1}{4} n)]}$ and $\varphi_0 = \sqrt{[(1/6)(2k - n)]}$.

For time part, we get

$$C_2 = \ln [\cosh b( t - t_0 )] \qquad \ldots(25)$$

$$\beta^2 = n - 2k + (2k/3) \tanh^2 b(t - t_0) \qquad \ldots(26)$$

[ where, $b^2 = k$ and $t_0$ is an integration constant.]

Since, $\varphi_0$ real, we have, $2k > n$. Also, as $t \to \infty$, $\beta^2 \geq 0$.

Thus, k is restricted as : $\tfrac{1}{2} n < k \leq \tfrac{3}{4} n$.

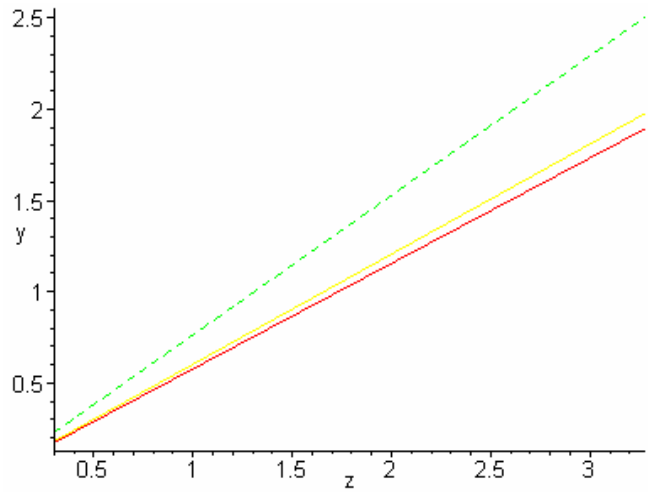

Fig-1: The variation of the scalar field y ≡ φ with respect to 'z' for different choices of the parameters ( n = 6 & k = 4  for red line; n = 7.5 & k = 4.84  for green line and n = 9 & k = 6.25  for yellow line ).

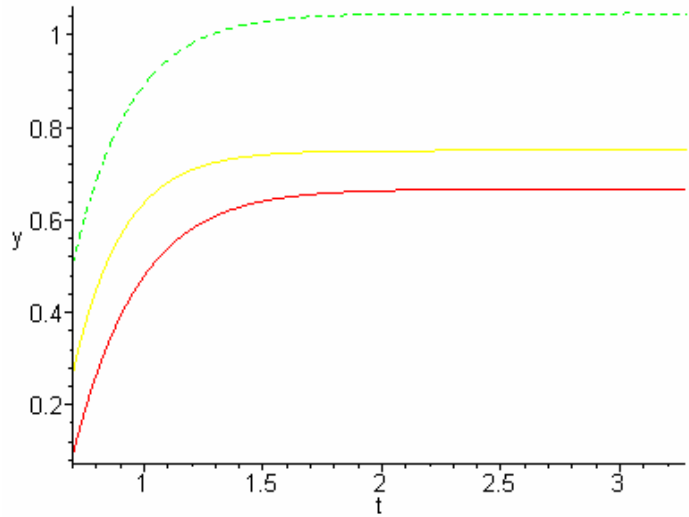

Fig-2:  The variation of  $\beta^2 \equiv$ y with respect to 't' for different choices of the parameters ( n = 6 & k = 4  for red line; n = 7.5 & k = 4.84  for green line and n = 9 & k = 6.25  for yellow line ).

Also we get an expression for potential V (φ) as

V (φ) = (1/12)( 2k – n ) exp(– $C_0$ z )   ....(27)

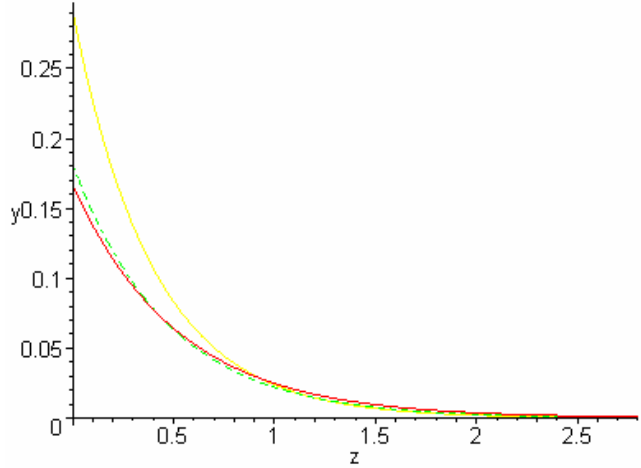

Fig-3: The variation of potential V (φ) ≡ y with respect to 'z' for different choices of the parameters ( n = 6 & k = 4 for red line; n = 7.5 & k = 4.84 for green line and n = 9 & k = 6.25 for yellow line ).

Thus the line element is given by

$ds^2$ = exp( $C_0$ z ) [$dt^2$ – $dz^2$ – cosh bt ($dx^2$ + $dy^2$) ]   ...(28)

[ neglecting integration constant $t_0$ ]

The energy density

ρ = $T_t^t$ = $T_x^x$ = $T_y^y$ = (1/6)( 2k – n ) exp(–$C_0$ z )   ....(29)

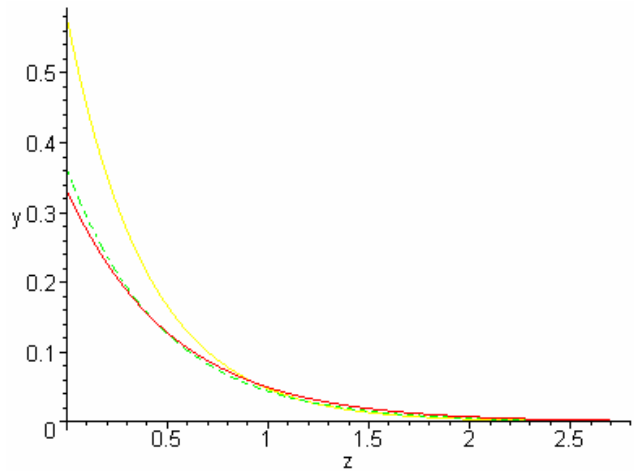

Fig-4: The variation of energy density ρ ≡ y with respect to 'z' for different choices of the parameters ( n = 6 & k = 4 for red line; n = 7.5 & k = 4.84 for green line and n = 9 & k = 6.25 for yellow line ).

The energy density ρ and potential energy of the field V both vanish at z → ∞.

It may be interesting to see if any particle horizon exists in our space time.
We now calculate the proper distance $S_H$ between z = 0 and z = ∞, measured along a space like curve running perpendicular to the wall (t, x, y = constant) as

$$S_H = \int_0^\infty \exp(½ A) dz = \int_0^\infty [\exp(½ C_0 z)] dx \qquad ...(30)$$

This distance diverges. Hence there is no horizon in the z- direction i.e. perpendicular to the wall. This is very similar to the result obtained by Wang [7] for a thick wall in Einstein's theory but at variance with Goetz's result [5].

Another area of interest may be the nature of the gravitational interaction of the wall. This can be studied by either the analysis of time like geodesics in the space time or the one of the acceleration of an observer who is at rest relative to the wall.
The z – component of the acceleration acting on a test particle in the gravitational field of domain wall is given by

$$A^r = V_{;\mu}{}^r V^\mu \qquad ....(31)$$

Since for a co moving particle, $V^\mu = \exp(-½ A) \delta^\mu{}_o$, then we find that

$$A^r = ½ C_0 \exp(-C_0 z) \qquad ....(32)$$

The above expression, being positive – definite, follows that an observer who wishes to remain stationary with respect to the wall must accelerate away from the wall. This means that the wall has an attractive influence on the test particle. This result is also conformity with Wang but differs from Goetz.

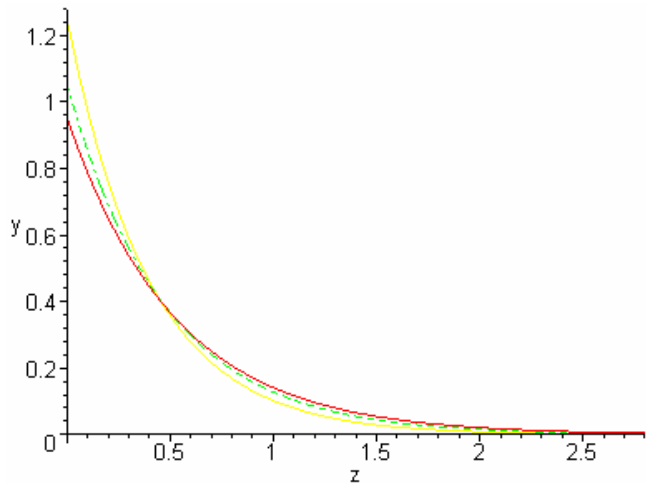

Fig-5: The variation of acceleration $A^r \equiv y$ with respect to 'z' for different choices of the parameters ( n = 6 & k = 4 for red line; n = 7.5 & k = 4.84 for green line and n = 9 & k = 6.25 for yellow line ).

## 5. Concluding Remarks:

Basically our motivation to investigate domain walls out side the general relativity scheme, specially in Lyra geometry as we believe scalar tensor theory can play an important role to understand of the early Universe, when topological defects may have existed.

To, summarize, we have found solutions of the coupled Einstein-scalar field equations describing thick domain wall in Lyra geometry with arbitrary scalar field distribution and vanishing energy momentum tensors far away from the wall.

In the hyper surfaces z = constant, the metric represents the three dimensional de-sitter spaces. The general expression for the three space volume is given by

$$\sqrt{|g_3|} = \sqrt{[\exp(3C_0 z) \cosh^2 bt\,]} \qquad \ldots(33)$$

Thus the temporal behavior would be

$$\sqrt{|g_3|} \sim \cosh bt \qquad \ldots(34)$$

One can see that the three space expands along z – direction.

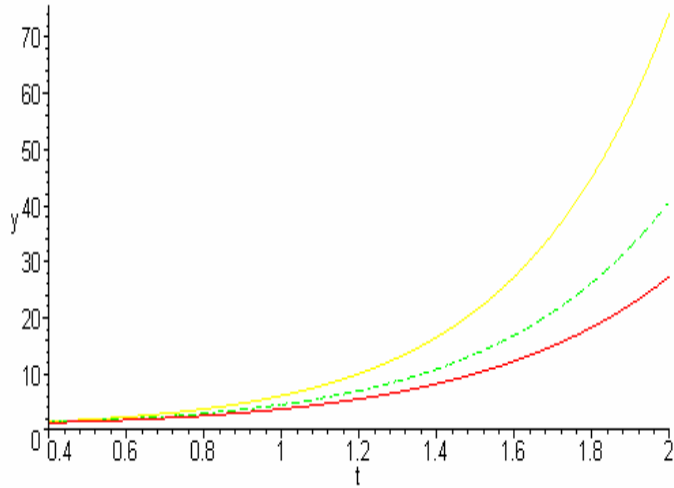

Fig-6: The temporal behavior of the three space $\sqrt{|g_3|} \equiv y$ with respect to 'z' for different choices of the parameters ( n = 6 & k = 4 for red line; n = 7.5 & k = 4.84 for green line and n = 9 & k = 6.25 for yellow line ).

Similar to the Goetz's wall, the energy density of the wall is time independent but in contrast to the Goetz's wall, the space time has no particle horizons and the gravitational field of the domain wall is attractive in nature.

However, in contrary to the Wang's wall, the energy density of the wall is time independent but in similar to the Wang's wall the space time has no particle horizons and the gravitational field of the domain wall is attractive in nature. Thus our thick domain wall in Lyra geometry exhibits peculiar features.

Though energy stress components are time independent, but displacement vector depends on time co-ordinate. The displacement vector becomes nonzero constant at large time i.e. it is surprising to note the displacement vector still exist after infinite time.

So concept of Lyra geometry is still exist even after the infinite times. As a future exercise, it will be interesting to study different properties of different topological defects within the frame work of Lyra geometry.

## Acknowledgements:

F.R is thankful to DST, Government of India for providing financial support.